\documentclass[aps,prl,showpacs,twocolumn,superscriptaddress]{revtex4}  

\usepackage{graphicx}  
\usepackage{dcolumn}   
\usepackage{bm}        
\usepackage{amssymb}   
\usepackage{epsfig}
\RequirePackage[usenames]{pstcol}

\def\met{{\mbox{$E\kern-0.57em\raise0.19ex\hbox{/}_{T}$}}}

\def\ifb{fb$^{-1}$}
\def\pp{$p\bar{p}$}

\def\lmet{$WH$$\rightarrow$$ \ell\kern-0.45em\raise0.19ex\hbox{/} \nu b\bar{b}$}

\def\hww{$H$$\rightarrow$$ W^+ W^-$}

\newcommand{\ttbar}{$t\overline{t}$}

\begin{document}

\leftline{\hspace*{13cm} FERMILAB-PUB-10-017-E}

\title{
Combination of Tevatron searches for the standard model Higgs
boson in the {\boldmath $W^+W^-$} decay mode}

\affiliation{Universidad de Buenos Aires, Buenos Aires, Argentina}
\affiliation{LAFEX, Centro Brasileiro de Pesquisas F{\'\i}sicas, Rio de Janeiro, Brazil}
\affiliation{Universidade do Estado do Rio de Janeiro, Rio de Janeiro, Brazil}
\affiliation{Universidade Federal do ABC, Santo Andr\'e, Brazil}
\affiliation{Instituto de F\'{\i}sica Te\'orica, Universidade Estadual Paulista, S\~ao Paulo, Brazil}
\affiliation{Institute of Particle Physics: McGill University, Montr\'{e}al, Qu\'{e}bec, Canada; Simon Fraser University, Burnaby, British Columbia, Canada; University of Toronto, Toronto, Ontario, Canada; and TRIUMF, Vancouver, British Columbia, Canada}
\affiliation{Simon Fraser University, Burnaby, British Columbia, Canada; and York University, Toronto, Ontario, Canada}
\affiliation{University of Science and Technology of China, Hefei, People's Republic of China}
\affiliation{Institute of Physics, Academia Sinica, Taipei, Taiwan, Republic of China}
\affiliation{Universidad de los Andes, Bogot\'{a}, Colombia}
\affiliation{Center for Particle Physics, Charles University, Faculty of Mathematics and Physics, Prague, Czech Republic}
\affiliation{Czech Technical University in Prague, Prague, Czech Republic}
\affiliation{Center for Particle Physics, Institute of Physics, Academy of Sciences of the Czech Republic, Prague, Czech Republic}
\affiliation{Universidad San Francisco de Quito, Quito, Ecuador}
\affiliation{Division of High Energy Physics, Department of Physics, University of Helsinki and Helsinki Institute of Physics, FIN-00014, Helsinki, Finland}
\affiliation{LPC, Universit\'e Blaise Pascal, CNRS/IN2P3, Clermont, France}
\affiliation{LPSC, Universit\'e Joseph Fourier Grenoble 1, CNRS/IN2P3, Institut National Polytechnique de Grenoble, Grenoble, France}
\affiliation{CPPM, Aix-Marseille Universit\'e, CNRS/IN2P3, Marseille, France}
\affiliation{LAL, Universit\'e Paris-Sud, CNRS/IN2P3, Orsay, France}
\affiliation{LPNHE, Universit\'es Paris VI and VII, CNRS/IN2P3, Paris, France}
\affiliation{CEA, Irfu, SPP, Saclay, France}
\affiliation{IPHC, Universit\'e de Strasbourg, CNRS/IN2P3, Strasbourg, France}
\affiliation{IPNL, Universit\'e Lyon 1, CNRS/IN2P3, Villeurbanne, France and Universit\'e de Lyon, Lyon, France}
\affiliation{III. Physikalisches Institut A, RWTH Aachen University, Aachen, Germany}
\affiliation{Physikalisches Institut, Universit{\"a}t Bonn, Bonn, Germany}
\affiliation{Physikalisches Institut, Universit{\"a}t Freiburg, Freiburg, Germany}
\affiliation{II. Physikalisches Institut, Georg-August-Universit{\"a}t G\"ottingen, G\"ottingen, Germany}
\affiliation{Institut f\"{u}r Experimentelle Kernphysik, Karlsruhe Institute of Technology, Karlsruhe, Germany}
\affiliation{Institut f{\"u}r Physik, Universit{\"a}t Mainz, Mainz, Germany}
\affiliation{Ludwig-Maximilians-Universit{\"a}t M{\"u}nchen, M{\"u}nchen, Germany}
\affiliation{Fachbereich Physik, University of Wuppertal, Wuppertal, Germany}
\affiliation{University of Athens, 157 71 Athens, Greece}
\affiliation{Panjab University, Chandigarh, India}
\affiliation{Delhi University, Delhi, India}
\affiliation{Tata Institute of Fundamental Research, Mumbai, India}
\affiliation{University College Dublin, Dublin, Ireland}
\affiliation{Istituto Nazionale di Fisica Nucleare Bologna, $^{ee}$University of Bologna, I-40127 Bologna, Italy}
\affiliation{Laboratori Nazionali di Frascati, Istituto Nazionale di Fisica Nucleare, I-00044 Frascati, Italy}
\affiliation{Istituto Nazionale di Fisica Nucleare, Sezione di Padova-Trento, $^{ff}$University of Padova, I-35131 Padova, Italy}
\affiliation{Istituto Nazionale di Fisica Nucleare Pisa, $^{gg}$University of Pisa, $^{hh}$University of Siena and $^{ii}$Scuola Normale Superiore, I-56127 Pisa, Italy}
\affiliation{Istituto Nazionale di Fisica Nucleare, Sezione di Roma 1, $^{jj}$Sapienza Universit\`{a} di Roma, I-00185 Roma, Italy}
\affiliation{Istituto Nazionale di Fisica Nucleare Trieste/Udine, I-34100 Trieste, $^{kk}$University of Trieste/Udine, I-33100 Udine, Italy}
\affiliation{Okayama University, Okayama 700-8530, Japan}
\affiliation{Osaka City University, Osaka 588, Japan}
\affiliation{University of Tsukuba, Tsukuba, Ibaraki 305, Japan}
\affiliation{Waseda University, Tokyo 169, Japan}
\affiliation{Center for High Energy Physics: Kyungpook National University, Daegu, Korea; Seoul National University, Seoul, Korea; Sungkyunkwan University, Suwon, Korea; Korea Institute of Science and Technology Information, Daejeon, Korea; Chonnam National University, Gwangju, Korea; Chonbuk National University, Jeonju, Korea}
\affiliation{Korea Detector Laboratory, Korea University, Seoul, Korea}
\affiliation{SungKyunKwan University, Suwon, Korea}
\affiliation{CINVESTAV, Mexico City, Mexico}
\affiliation{FOM-Institute NIKHEF and University of Amsterdam/NIKHEF, Amsterdam, The Netherlands}
\affiliation{Radboud University Nijmegen/NIKHEF, Nijmegen, The Netherlands}
\affiliation{Joint Institute for Nuclear Research, Dubna, Russia}
\affiliation{Institution for Theoretical and Experimental Physics, Moscow, Russia}
\affiliation{Moscow State University, Moscow, Russia}
\affiliation{Institute for High Energy Physics, Protvino, Russia}
\affiliation{Petersburg Nuclear Physics Institute, St. Petersburg, Russia}
\affiliation{Comenius University, 842 48 Bratislava, Slovakia; Institute of Experimental Physics, 040 01 Kosice, Slovakia}
\affiliation{Institut de Fisica d'Altes Energies, Universitat Autonoma de Barcelona, E-08193, Bellaterra (Barcelona), Spain}
\affiliation{Centro de Investigaciones Energeticas Medioambientales y Tecnologicas, E-28040 Madrid, Spain}
\affiliation{Instituto de Fisica de Cantabria, CSIC-University of Cantabria, 39005 Santander, Spain}
\affiliation{Stockholm University, Stockholm, Sweden, and Uppsala University, Uppsala, Sweden}
\affiliation{University of Geneva, CH-1211 Geneva 4, Switzerland}
\affiliation{Glasgow University, Glasgow G12 8QQ, United Kingdom}
\affiliation{Lancaster University, Lancaster LA1 4YB, United Kingdom}
\affiliation{University of Liverpool, Liverpool L69 7ZE, United Kingdom}
\affiliation{Imperial College London, London SW7 2AZ, United Kingdom}
\affiliation{University College London, London WC1E 6BT, United Kingdom}
\affiliation{The University of Manchester, Manchester M13 9PL, United Kingdom}
\affiliation{University of Oxford, Oxford OX1 3RH, United Kingdom}
\affiliation{University of Arizona, Tucson, Arizona 85721, USA}
\affiliation{Ernest Orlando Lawrence Berkeley National Laboratory, Berkeley, California 94720, USA}
\affiliation{University of California, Davis, Davis, California 95616, USA}
\affiliation{University of California, San Diego, La Jolla, California 92093, USA}
\affiliation{University of California, Los Angeles, Los Angeles, California 90024, USA}
\affiliation{University of California, Riverside, Riverside, California 92521, USA}
\affiliation{University of California, Santa Barbara, Santa Barbara, California 93106, USA}
\affiliation{Yale University, New Haven, Connecticut 06520, USA}
\affiliation{University of Florida, Gainesville, Florida 32611, USA}
\affiliation{Florida State University, Tallahassee, Florida 32306, USA}
\affiliation{Argonne National Laboratory, Argonne, Illinois 60439, USA}
\affiliation{Fermi National Accelerator Laboratory, Batavia, Illinois 60510, USA}
\affiliation{Enrico Fermi Institute, University of Chicago, Chicago, Illinois 60637, USA}
\affiliation{University of Illinois at Chicago, Chicago, Illinois 60607, USA}
\affiliation{Northern Illinois University, DeKalb, Illinois 60115, USA}
\affiliation{Northwestern University, Evanston, Illinois 60208, USA}
\affiliation{University of Illinois, Urbana, Illinois 61801, USA}
\affiliation{Indiana University, Bloomington, Indiana 47405, USA}
\affiliation{Purdue University Calumet, Hammond, Indiana 46323, USA}
\affiliation{University of Notre Dame, Notre Dame, Indiana 46556, USA}
\affiliation{Purdue University, West Lafayette, Indiana 47907, USA}
\affiliation{Iowa State University, Ames, Iowa 50011, USA}
\affiliation{University of Kansas, Lawrence, Kansas 66045, USA}
\affiliation{Kansas State University, Manhattan, Kansas 66506, USA}
\affiliation{Louisiana Tech University, Ruston, Louisiana 71272, USA}
\affiliation{The Johns Hopkins University, Baltimore, Maryland 21218, USA}
\affiliation{University of Maryland, College Park, Maryland 20742, USA}
\affiliation{Boston University, Boston, Massachusetts 02215, USA}
\affiliation{Northeastern University, Boston, Massachusetts 02115, USA}
\affiliation{Harvard University, Cambridge, Massachusetts 02138, USA}
\affiliation{Massachusetts Institute of Technology, Cambridge, Massachusetts 02139, USA}
\affiliation{Tufts University, Medford, Massachusetts 02155, USA}
\affiliation{Brandeis University, Waltham, Massachusetts 02254, USA}
\affiliation{University of Michigan, Ann Arbor, Michigan 48109, USA}
\affiliation{Wayne State University, Detroit, Michigan 48201, USA}
\affiliation{Michigan State University, East Lansing, Michigan 48824, USA}
\affiliation{University of Mississippi, University, Mississippi 38677, USA}
\affiliation{University of Nebraska, Lincoln, Nebraska 68588, USA}
\affiliation{Rutgers University, Piscataway, New Jersey 08855, USA}
\affiliation{Princeton University, Princeton, New Jersey 08544, USA}
\affiliation{University of New Mexico, Albuquerque, New Mexico 87131, USA}
\affiliation{State University of New York, Buffalo, New York 14260, USA}
\affiliation{Columbia University, New York, New York 10027, USA}
\affiliation{The Rockefeller University, New York, New York 10021, USA}
\affiliation{University of Rochester, Rochester, New York 14627, USA}
\affiliation{State University of New York, Stony Brook, New York 11794, USA}
\affiliation{Brookhaven National Laboratory, Upton, New York 11973, USA}
\affiliation{Duke University, Durham, North Carolina 27708, USA}
\affiliation{The Ohio State University, Columbus, Ohio 43210, USA}
\affiliation{Langston University, Langston, Oklahoma 73050, USA}
\affiliation{University of Oklahoma, Norman, Oklahoma 73019, USA}
\affiliation{Oklahoma State University, Stillwater, Oklahoma 74078, USA}
\affiliation{University of Pennsylvania, Philadelphia, Pennsylvania 19104, USA}
\affiliation{Carnegie Mellon University, Pittsburgh, Pennsylvania 15213, USA}
\affiliation{University of Pittsburgh, Pittsburgh, Pennsylvania 15260, USA}
\affiliation{Brown University, Providence, Rhode Island 02912, USA}
\affiliation{University of Texas, Arlington, Texas 76019, USA}
\affiliation{Texas A\&M University, College Station, Texas 77843, USA}
\affiliation{Southern Methodist University, Dallas, Texas 75275, USA}
\affiliation{Rice University, Houston, Texas 77005, USA}
\affiliation{Baylor University, Waco, Texas 76798, USA}
\affiliation{University of Virginia, Charlottesville, Virginia 22901, USA}
\affiliation{University of Washington, Seattle, Washington 98195, USA}
\affiliation{University of Wisconsin, Madison, Wisconsin 53706, USA}

\author{T.~Aaltonen$^{\dag}$} \affiliation{Division of High Energy Physics, Department of Physics, University of Helsinki and Helsinki Institute of Physics, FIN-00014, Helsinki, Finland}
\author{V.M.~Abazov$^{\ddag}$} \affiliation{Joint Institute for Nuclear Research, Dubna, Russia}
\author{B.~Abbott$^{\ddag}$} \affiliation{University of Oklahoma, Norman, Oklahoma 73019, USA}
\author{M.~Abolins$^{\ddag}$} \affiliation{Michigan State University, East Lansing, Michigan 48824, USA}
\author{B.S.~Acharya$^{\ddag}$} \affiliation{Tata Institute of Fundamental Research, Mumbai, India}
\author{M.~Adams$^{\ddag}$} \affiliation{University of Illinois at Chicago, Chicago, Illinois 60607, USA}
\author{T.~Adams$^{\ddag}$} \affiliation{Florida State University, Tallahassee, Florida 32306, USA}
\author{J.~Adelman$^{\dag}$} \affiliation{Enrico Fermi Institute, University of Chicago, Chicago, Illinois 60637, USA}
\author{E.~Aguilo$^{\ddag}$} \affiliation{Simon Fraser University, Burnaby, British Columbia, Canada; and York University, Toronto, Ontario, Canada}
\author{G.D.~Alexeev$^{\ddag}$} \affiliation{Joint Institute for Nuclear Research, Dubna, Russia}
% ...

\collaboration{The CDF$^\dag$ and D0$^\ddag$ Collaborations} \noaffiliation

\date{\today}

\begin{abstract}
~\\ We combine searches by the CDF and D0 collaborations for a Higgs
boson decaying to $W^+W^-$.  The data correspond to an integrated
total luminosity of 4.8 (CDF) and 5.4 (D0) fb$^{-1}$ of $p{\bar{p}}$ collisions
at $\sqrt{s}=1.96$~TeV at the Fermilab Tevatron collider.  No excess
is observed above background expectation, and resulting limits on
Higgs boson production exclude a standard-model Higgs boson in the
mass range 162--166 GeV at the 95\% C.L.
\end{abstract}
\maketitle

\clearpage
\newpage

Finding the last unobserved fundamental particle in the standard model
(SM), the Higgs boson, is a major goal of particle physics, and the
search for its existence is a central component of Fermilab's Tevatron
program.  Direct searches at the CERN LEP collider have set a limit on
the Higgs boson mass of $m_H$$>$$114.4$~GeV at the 95\%
C.L.~\cite{sm-lep}.  Combining this limit with precision electroweak
measurements constrains the mass of the SM Higgs boson to be less than
186~GeV at the 95\% C.L.~\cite{elweak}.  The favored mass range
therefore places the SM Higgs boson within the reach of the
experiments at the Fermilab Tevatron collider.

In this Letter, we combine searches for Higgs bosons ($H$) decaying to
$W^+W^-$ performed by the CDF and D0
Collaborations~\cite{cdfHWW,dzHWW}. These searches are particularly
sensitive to a Higgs boson with mass 130$<$$m_{H}$$<$200~GeV.  The
data analyzed correspond to integrated luminosities of 4.8~\ifb\ and
5.4~\ifb\ collected with the CDF and D0 detectors, respectively.  We
use all significant production modes, namely, gluon-gluon fusion
($gg$$\rightarrow$$ H$), associated production
($q{\bar{q}}$$\rightarrow$$ WH$ or $ZH$), and vector boson fusion
($q{\bar{q}}$$\rightarrow$$ q{\bar{q}}H$, where the quarks radiate
weak gauge bosons that fuse to form the $H$, and is referred to as
VBF).

The event selections used in the CDF and D0 analyses are similar.
Both collaborations select events with large missing transverse energy
and two oppositely charged, isolated leptons, targeting the \hww\
signal in which both $W$ bosons decay leptonically.  The D0 selection
classifies events in three channels, $e^+e^-$, $e^\pm \mu^\mp$, and
$\mu^+\mu^-$.  The CDF selection separates opposite-sign di-lepton
candidate events into five non-overlapping channels, classifying
events by their jet multiplicity (0, 1, or $\ge$ 2); the 0- and 1-jet
channels are further divided depending on whether one or both leptons
are in the central part of the detector. In addition, CDF searches for
Higgs boson events containing same-sign lepton pairs, mainly produced
in $WH$ and $ZH$ associated production.

The presence of neutrinos in the final state prevents full
reconstruction of the Higgs boson mass.  Other variables are used to
search for a signal in the presence of appreciable background.  For
example, the azimuthal angle between the leptons in signal events is
smaller on average than that in background events due to the scalar
nature of the Higgs boson and the parity violation in $W^\pm$ decays.
The missing transverse energy is larger, and the total transverse
energy of the jets is smaller, in signal events than in background
events.  The final discriminants are binned neural-network outputs
based on several kinematic input variables~\cite{cdfHWW,dzHWW}. A
dedicated network is trained for each Higgs boson mass tested.  For
CDF, the inputs include likelihoods constructed from matrix-element
probabilities.  Compared with earlier Tevatron $H$$\rightarrow$$
W^+W^-$ analyses, the new analyses use larger data samples, include
all significant signal production mechanisms, and have undergone additional
improvements in search sensitivity.

The Higgs boson signals are simulated with {\sc pythia}~\cite{pythia},
using CTEQ5L~\cite{cteq5l} (CDF) and CTEQ6L1~\cite{cteq6l} (D0) parton
distribution functions (PDF) at leading order (LO).  We normalize our
predictions for the Higgs boson signals to the most recent higher-order
perturbative QCD calculations available.
References~\cite{anastasiou,grazzinideflorian} and references therein
provide the steps used to calculate the $gg$$\rightarrow$$ H$ cross
section.  The MSTW 2008 next-to-next-to-leading order (NNLO) PDF
set~\cite{mstw2008} is used to predict the $gg$$\rightarrow$$ H$
production cross section.  The calculations of associated production
and VBF cross sections are described in
Refs.~\cite{vtheory,vhtheory,vbftheory}.  The branching fractions for
the Higgs boson decays are obtained from {\sc hdecay}~\cite{hdecay}.
After all selections, the total number of expected Higgs boson events
is approximately 30 per experiment for $m_H$$=$165 GeV, which
corresponds to the region of greatest sensitivity.

Both experiments determine the multijet background by studying control
samples, which are then extrapolated into the signal regions.  For
CDF, backgrounds from SM $WW$, $WZ$, $ZZ$, $W\gamma$,
Drell-Yan, and $t{\bar{t}}$ production are generated using the {\sc
pythia}, {\sc mc@nlo}~\cite{MC@NLO}, and the {\sc ub/eb}~\cite{Baur}
programs.  Backgrounds from $W+$jets processes, including (for CDF)
semileptonic diboson events, single top, and semileptonic $t{\bar{t}}$
events, are modeled using $W+$jets data events and a measurement
of the rate at which jets are misidentified as leptons.  For D0, these
backgrounds are generated using {\sc pythia}, {\sc
alpgen}~\cite{alpgen}, and {\sc comphep}~\cite{comphep}, with {\sc
pythia} providing parton-showering and hadronization for all
generators.

The diboson backgrounds are normalized using next-to-leading order (NLO)
calculations from {\sc mcfm}~\cite{mcfm}.
The $t{\bar{t}}$ and single top production cross sections are taken
from Refs.~\cite{mochuwer,mrst06} and Ref.~\cite{kidonakis_st}
respectively.  NNLO calculations~\cite{hamberg} are used by both the CDF
and D0 Collaborations for the Drell-Yan background, and by D0 for the
inclusive $W$/$Z$ processes. Other backgrounds are normalized to experimental data.
Both collaborations use NLO simulations and data control samples to
improve the modeling of differential distributions.
Systematic uncertainties on the rates of the expected signal and the
expected backgrounds, as well as on the shapes of the final
discriminants, are included.  More details are given in
Refs.~\cite{cdfHWW,dzHWW}.

We perform the combination twice, using Bayesian and modified
frequentist approaches in turn. We check the consistency of the
results to verify that the final result does not depend on the details
of the statistical formulation.  Both combinations test signal mass
hypotheses in 5~GeV steps for values of $m_H$ between 130 and 200 GeV,
i.e., the mass range in which \hww\ is the dominant decay mode.
These two combinations give similar results (the limits agree within 5\%).
Both methods use the distributions of the
final discriminants, and not just the total event counts passing selection requirements.

Both statistical procedures form, for a given Higgs boson mass, a
combined likelihood (including priors on systematic uncertainties,
$\pi({\vec{\theta}})$) based on the product of likelihoods for the
individual channels, each of which in turn is a product over histogram
bins:
\begin{equation}
{\cal{L}}(R,{\vec{s}},{\vec{b}}|{\vec{n}},{\vec{\theta}})\times\pi({\vec{\theta}})
= \prod_{i=1}^{N_{\rm C}}\prod_{j=1}^{N_{\rm bins}} \mu_{ij}^{n_{ij}} \frac{e^{-\mu_{ij}}}{n_{ij}!}
\times\prod_{k=1}^{n_{\rm sys}}e^{-\theta_k^2/2}
\end{equation}
\noindent where the first product is over the number of channels
($N_{\rm C}$), and the second product is over histogram bins
containing $n_{ij}$ events, binned in ranges of the final
neural-network discriminants used for the individual analyses.  The
predictions for the bin contents are $\mu_{ij} =R \times
s_{ij}({\vec{\theta}}) + b_{ij}({\vec{\theta}})$ for channel $i$ and
histogram bin $j$, where $s_{ij}$ and $b_{ij}$ represent the expected
SM signal and background in the bin, and $R$ is a scaling factor
applied to the signal.  By scaling all signal contributions by the
same factor we make the assumption that the relative contributions of
the different processes at each $m_H$ are as given by the SM.
Systematic uncertainties are parameterized by the dependence of
$s_{ij}$ and $b_{ij}$ on ${\vec\theta}$.  Each of the $n_{\rm sys}$
components of ${\vec\theta}$, $\theta_k$, corresponds to a single
independent source of systematic uncertainty scaled by its standard deviation,
and each parameter may
have an impact on several sources of signal and background in
different channels, thus accounting for correlations.

In the Bayesian method we assume a uniform prior in
the signal yield.  Gaussian
priors are assumed for the $\theta_k$, truncated so that no prediction
is negative.  The posterior density function is then integrated over
the $\theta_k$ (including correlations) and a 95\% C.L. upper limit on
$R$, $R_{\mathrm{lim}}$, satisfies
\begin{equation}
\frac{
\int_0^{R_{\mathrm{lim}}}\int{\cal{L}}(R,{\vec{s}},{\vec{b}}|{\vec{n}},{\vec{\theta}})\pi({\vec{\theta}})d{\vec{\theta}}dR}{
\int_0^{\infty}\int{\cal{L}}(R,{\vec{s}},{\vec{b}}|{\vec{n}},{\vec{\theta}})\pi({\vec{\theta}})d{\vec{\theta}}dR}
=0.95.
\end{equation}

The modified frequentist technique uses the statistical variable
CL$_{\rm s}$, defined in Ref.~\cite{pdgstats}, to test hypotheses
which correspond to the presence or absence of Higgs boson signals.
The test statistic is the log-likelihood ratio ${\rm
LLR}=-2\ln\frac{p({\mathrm{data}}|{\mathrm{s+b}})}{p({\mathrm{data}}|{\mathrm{b}})}$, where
$p({\mathrm{data}}|{\mathrm{s+b}})$ and $p({\mathrm{data}}|{\mathrm{b}})$ are the
probabilities that the data are drawn from the s+b and b-only
hypotheses respectively.
The probabilities $p$ are computed using the best-fit values of the
parameters $\theta_k$, separately for each of the two
hypotheses~\cite{pflh}.  The use of these fits extends the procedure
used at LEP~\cite{pdgstats}, improving the sensitivity when the
expected signals are small and the uncertainties on the backgrounds
are large.  Two $p$-values are computed: CL$_{\mathrm{b}} =
p($LLR$\ge$ LLR$_{\mathrm{obs}} | {\mathrm{b}})$ and CL$_{\mathrm{s+b}} =
p($LLR$\ge$ LLR$_{\mathrm{obs}} | {\mathrm{s+b}})$, where LLR$_{\mathrm{obs}}$ is
the value of the test statistic computed for the data.  The ratio
CL$_{\mathrm{s}}={\mathrm{CL}}_{\mathrm{s+b}}/{\mathrm{CL}}_{\mathrm{b}}$
is used to define confidence intervals and is chosen to reduce the
potential for excluding a signal for which there is insufficient
sensitivity.  If CL$_{\mathrm{s}}$$<$$0.05$ for a particular choice of
s+b, that hypothesis is excluded at the 95\% C.L.  Systematic
uncertainties are included by fluctuating the predictions for $s_{ij}$
and $b_{ij}$ when generating the pseudoexperiments used to compute
CL$_{\mathrm{s+b}}$ and CL$_{\mathrm{b}}$.

Though many sources of systematic uncertainty differ between the
experiments and analyses, all appropriate correlations are taken into
account in the combined limits.  The dominant systematic uncertainties
arise from cross section calculations for the signals and the
backgrounds; these are correlated between the experiments.
Variations of the parton distribution functions and the
renormalization and factorization scales give rise to uncertainties of
11\% for the gluon-gluon fusion process~\cite{ana-dis}, 5\% for
associated $WH$ and $ZH$ production~\cite{vtheory,vhtheory}, and 10\% for
VBF~\cite{vtheory,vbftheory}.  CDF, which uses analyses separated in
jet multiplicity bins, applies a channel (jet bin) dependent
uncertainty of 7\% to 70\% and a gluon PDF uncertainty of 8\% on $gg
$$\rightarrow$$ H$, following the treatment discussed in
Ref.~\cite{ana-dis}.  For the gluon-gluon fusion signal process, we
study the effects on the acceptance and the kinematics of scale
variations, gluon PDF variations, and the differences between
next-to-next-to-leading log calculations and the generators used for
the central predictions, using the {\sc fehip} and {\sc hnnlo}
programs~\cite{anastasiou,grazzini}.  We find additional uncertainties
of 5\% to 10\%. The primary background, $W^+W^-$ production, has a
cross section uncertainty of 7\% and a similar study of the acceptance
and kinematics finds additional uncertainties of approximately 1\% to
5\%. The systematic uncertainties on $WZ$, $ZZ$, \ttbar, single top
production, and Drell-Yan production range from 7\% to 10\%.  The
uncertainties on the multijet background are uncorrelated between the
experiments and range from 2\% to 15\%.  The uncertainties on the
yields of $W+$jets and $W\gamma$(+jets) range from 7\% to 30\%, but
these have small effects on the results because the rates of these
backgrounds are low.  Because the methods of estimating the $W+$ jets
and $W\gamma$(+jets) backgrounds differ between CDF and D0, we assume
there is no correlation between these rates.  The uncertainties on the
lepton identification and the trigger efficiencies are uncorrelated
between the experiments and range from 2\% to 6\%.  The uncertainty on
the integrated luminosity of 6\% is taken to be correlated between the
signal and the Monte-Carlo-based background predictions, and partially
correlated between the experiments, via the 4\% uncertainty on the
inelastic $p{\bar{p}}$ cross section~\cite{inelppbar}.  Additional
details related to the treatment of systematic uncertainties are
given in Refs.~\cite{cdfHWW,dzHWW}.
As bin by bin uncertainties arising from the statistical uncertainty
in the Monte Carlo (and in some cases data) samples were shown to
affect the observed and expected limits by less than 1\%, they are
neglected.

To better visualize the impact of the data events, we combine the
 histograms of the final discriminants, adding the contents of bins
 with similar $s/b$ ratios, so as not to dilute the impact of
 highly-sensitive bins with those with less discriminating power.
 Figure~\ref{fig:bgsub} shows the signal expectation and the data with
 the background subtracted, as a function of the $s/b$ ratio of the
 collected bins.  The background model has been fit to the data, and
 the uncertainties on the background are those after the systematic
 uncertainties have been constrained by the fit.  No excess of
 candidate events in the highest $s/b$ bins relative to the background
 expectation is observed.

 \begin{figure}[htb] \begin{centering}
 \includegraphics[width=0.8\columnwidth]{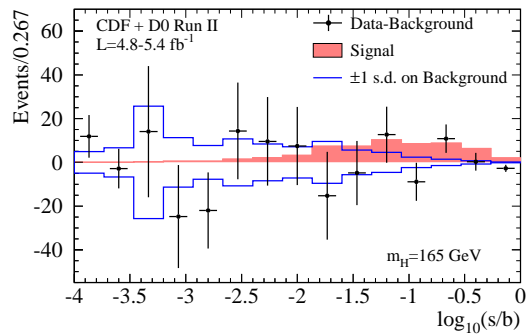} \caption{
 \label{fig:bgsub} (color online). Background-subtracted data distributions for the
 discriminant histograms, summed for bins with similar $s/b$, for the
 $m_H=165$~GeV combined search.  The background has been fit to the
 data under the b-only hypothesis, and the uncertainty on the
 background is the post-fit systematic uncertainty.  The signal, which
 is normalized to the SM expectation, is shown with a filled
 histogram.  The uncertainties shown on the background-subtracted data points
 are the square roots of the post-fit background predictions in
 each bin, representing the expected statistical uncertainty on the
 data.  } \end{centering} \end{figure}

Before extracting the combined limits, we study the LLR distributions
for the s+b and b-only hypotheses, shown in
Fig.~\ref{fig:comboLLR} as functions of $m_H$.  The separation between
the median LLR$_{\mathrm{b}}$ and LLR$_{\mathrm{s+b}}$ divided by the
widths is a measure of the discriminating power of the search.  The
value of LLR$_{\mathrm{obs}}$ relative to the expected s+b and
b-only distributions indicates whether the observed data are more
consistent with the presence of signal, or not.  No significant excess
of data above the background expectation is seen for any value of
$m_H$.  Because the same data events are used to construct the
observed LLR at each $m_H$ tested, the LLR values are highly
correlated from one $m_H$ to the next.  This also applies to
Figs.~\ref{fig:comboRatio} and~\ref{fig:comboLLR-2} described below.

 \begin{figure}[htb] \begin{centering}
 \includegraphics[width=7.4cm]{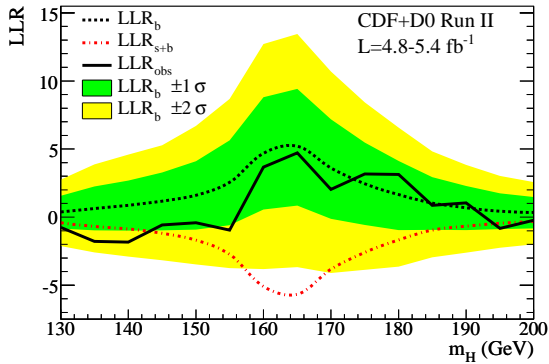} \caption{
 \label{fig:comboLLR} (color online). Distributions of LLR as functions of the Higgs
 boson mass.  We display the median values of the LLR distribution for
 the b-only hypothesis (LLR$_{\mathrm{b}}$), the s+b hypothesis
 (LLR$_{\mathrm{s+b}}$), and for the data (LLR$_{\mathrm{obs}}$).
 The shaded bands indicate the 68\% and 95\% probability regions in
 which the LLR is expected to fluctuate, in the absence of signal.  }
 \end{centering} \end{figure}

We extract limits on SM Higgs boson production in \pp~collisions at
$\sqrt{s}=1.96$~TeV in the $m_H=130$-$200$~GeV mass range.  We present
our results in terms of $R_{\mathrm{lim}}$, the ratio of the limits
obtained to the rate predicted by the SM, as a function of the Higgs
boson mass.  We assume the production fractions for $WH$, $ZH$,
$gg$$\rightarrow$$ H$, and VBF, and the Higgs boson decay branching
fractions, are those predicted by the SM.  A value of
$R_{\mathrm{lim}}$ less than or equal to one indicates a Higgs boson
mass that is excluded at the 95\% C.L.

\begin{figure}[t]
\begin{centering}
\includegraphics[width=7.4cm]{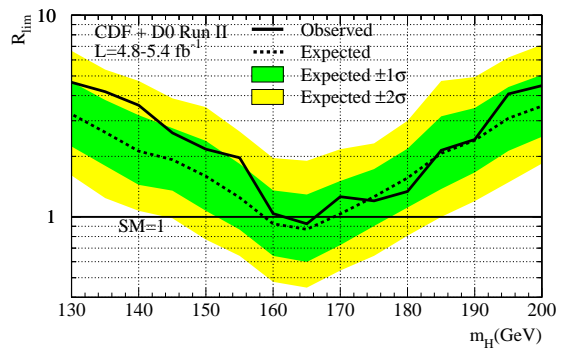}
\caption{
\label{fig:comboRatio} (color online).
Observed and expected (median, for the background-only hypothesis)
95\% C.L. upper limits on SM Higgs boson production.  The shaded bands
indicate the 68\% and 95\% probability regions in which $R_{\rm{lim}}$
is expected to fluctuate, in the absence of signal.  The limits
displayed in this figure are obtained with the Bayesian calculation.
}
\end{centering}
\end{figure}

 \begin{figure}[h]
 \begin{centering}
 \includegraphics[width=7.4cm]{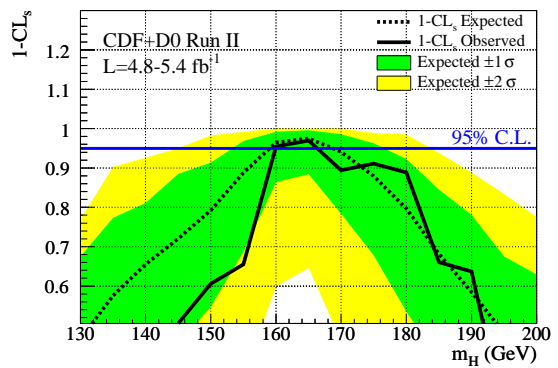}
 \caption{
 \label{fig:comboLLR-2} (color online).
 Distribution of 1-CL$_{\mathrm{s}}$ as a function of the Higgs boson mass
obtained with the CL$_{\mathrm{s}}$ method. The shaded bands indicate
the 68\% and 95\% probability regions in which the LLR is expected to
fluctuate, in the absence of signal. } \end{centering} \end{figure}

\begin{table*}[bt]
\caption{\label{tab:ratios} Ratios, $R_{\rm lim}$, of the median expected and observed 95\% C.L. limits
to the SM cross section for the combination of CDF and D0 analyses as
a function of the Higgs boson mass in GeV, obtained with the Bayesian
(upper) and the CL$_{\mathrm{s}}$ (lower)
methods.}\begin{ruledtabular}
\begin{tabular}{lccccccccccccccc}
Bayesian           &  130 &  135 &  140 &  145 &  150     &  155 &  160 &  165  &  170 &  175  &   180 &  185&  190&  195   & 200     \\ \hline
Expected           & 3.24 & 2.63 & 2.12 & 1.92 & 1.59     & 1.25 & 0.92 & 0.87  & 1.04 & 1.26  &  1.56 & 2.07 & 2.40 & 3.09 & 3.55    \\
Observed           & 4.65 & 4.18 & 3.58 & 2.61 & 2.17     & 1.96 & 1.04 & 0.93  & 1.26 & 1.20  &  1.34 & 2.14 & 2.42 & 4.07 & 4.47    \\
\hline	                                                                      	                  				 
CL${\rm_s}$        &  130 &  135 &  140 &  145 &  150     &  155 &  160 &  165  &  170 &  175  &   180 &  185&  190&  195   & 200     \\ \hline
Expected           & 3.26 & 2.52 & 2.18 & 1.87 & 1.53     & 1.24 & 0.89 & 0.84  & 1.06 & 1.28  &  1.56 & 2.07 & 2.46 & 3.17 & 3.62    \\
Observed           & 4.49 & 4.06 & 3.45 & 2.49 & 2.12     & 1.84 & 0.98 & 0.89  & 1.21 & 1.18  &  1.31 & 2.15 & 2.36 & 4.10 & 4.35    \\
\end{tabular}
\end{ruledtabular}
\end{table*}

The ratios of the expected and observed limits to the SM cross section
are shown in Fig.~\ref{fig:comboRatio} as a function of $m_H$.  The
observed and median expected ratios are listed in
Table~\ref{tab:ratios}, with observed (expected) values for the
Bayesian method of 1.04 (0.92) at $m_{H}=160$~GeV, 0.93 (0.87) at
$m_{H}=165$~GeV, and 1.26 (1.04) at $m_{H}=170$~GeV.  We use piecewise
linear interpolations to display the combination results in
Figs.~\ref{fig:comboLLR}--\ref{fig:comboLLR-2}, and to quote the
observed and expected excluded mass ranges.
We exclude the SM Higgs boson in the mass range 162 to 166 GeV.
The Bayesian calculation, chosen {\it a priori}, was used for this
exclusion.  The corresponding expected exclusion, from 159 to 169 GeV,
encompasses the observed exclusion.  The CL$_{\mathrm{s}}$ calculation
yields similar results, as shown in Fig.~\ref{fig:comboLLR-2}. The
1-CL$_{\mathrm{s}}$ distribution, which can be directly interpreted as
the level of exclusion of our search, is displayed as a function of
the Higgs boson mass. For instance, our expected limit shows that in
the absence of
signal the median 1-CL$_{\mathrm{s}}$ value with which
we expect to exclude a SM Higgs boson of mass 165~GeV is 97\%.

In summary, we present the first combined Tevatron search for
the SM Higgs boson using the $H$$\rightarrow$$ W^+W^-$ decay mode.  No
significant excess of candidates is found above the background
expectation for 130$<$$m_H$$<$200~GeV. We exclude the mass range from
162 to 166 GeV at the 95\% C.L.  This is the first direct constraint
on the mass of the Higgs boson beyond that obtained at LEP.

\begin{center}
{\bf Acknowledgements}
\end{center}

We thank the Fermilab staff and the technical staffs of the
participating institutions for their vital contributions. 
This work was supported by  
DOE and NSF (USA),
CONICET and UBACyT (Argentina), 
CNPq, FAPERJ, FAPESP and FUNDUNESP (Brazil),
CRC Program, CFI, NSERC and WestGrid Project (Canada),
CAS and CNSF (China),
Colciencias (Colombia),
MSMT and GACR (Czech Republic),
Academy of Finland (Finland),
CEA and CNRS/IN2P3 (France),
BMBF and DFG (Germany),
Ministry of Education, Culture, Sports, Science and Technology (Japan), 
World Class University Program, National Research Foundation (Korea),
KRF and KOSEF (Korea),
DAE and DST (India),
SFI (Ireland),
INFN (Italy),
CONACyT (Mexico),
NSC(Republic of China),
FASI, Rosatom and RFBR (Russia),
Slovak R\&D Agency (Slovakia), 
Ministerio de Ciencia e Innovaci\'{o}n, and Programa Consolider-Ingenio 2010 (Spain),
The Swedish Research Council (Sweden),
Swiss National Science Foundation (Switzerland), 
FOM (The Netherlands),
STFC and the Royal Society (UK),
and the A.P. Sloan Foundation (USA).

\begin{small}
With visitors to CDF from 
$^{a\dag}$University of Massachusetts Amherst, Amherst, Massachusetts 01003,
$^{b\dag}$Universiteit Antwerpen, B-2610 Antwerp, Belgium, 
$^{c\dag}$University of Bristol, Bristol BS8 1TL, United Kingdom,
$^{d\dag}$Chinese Academy of Sciences, Beijing 100864, China, 
$^{e\dag}$Istituto Nazionale di Fisica Nucleare, Sezione di Cagliari, 09042 Monserrato (Cagliari), Italy,
$^{f\dag}$University of California Irvine, Irvine, CA  92697, 
$^{g\dag}$University of California Santa Cruz, Santa Cruz, CA  95064, 
$^{h\dag}$Cornell University, Ithaca, NY  14853, 
$^{i\dag}$University of Cyprus, Nicosia CY-1678, Cyprus, 
$^{j\dag}$University College Dublin, Dublin 4, Ireland,
$^{k\dag}$University of Edinburgh, Edinburgh EH9 3JZ, United Kingdom, 
$^{l\dag}$University of Fukui, Fukui City, Fukui Prefecture, Japan 910-0017,
$^{m\dag}$Kinki University, Higashi-Osaka City, Japan 577-8502,
$^{n\dag}$Universidad Iberoamericana, Mexico D.F., Mexico,
$^{o\dag}$University of Iowa, Iowa City, IA  52242,
$^{p\dag}$Iowa State University, Ames, IA 50011,   
$^{q\dag}$Kansas State University, Manhattan, KS 66506,
$^{r\dag}$Queen Mary, University of London, London, E1 4NS, UK,
$^{s\dag}$University of Manchester, Manchester M13 9PL, UK,
$^{t\dag}$Muons, Inc., Batavia, IL 60510, 
$^{u\dag}$Nagasaki Institute of Applied Science, Nagasaki, Japan, 
$^{v\dag}$University of Notre Dame, Notre Dame, IN 46556,
$^{w\dag}$Obninsk State University, Obninsk, Russia,
$^{x\dag}$University de Oviedo, E-33007 Oviedo, Spain, 
$^{y\dag}$Texas Tech University, Lubbock, TX  79609, 
$^{z\dag}$IFIC(CSIC-Universitat de Valencia), 56071 Valencia, Spain,
$^{aa\dag}$Universidad Tecnica Federico Santa Maria, 110v Valparaiso, Chile,
$^{bb\dag}$University of Virginia, Charlottesville, VA  22906,
$^{cc\dag}$Bergische Universit\"at Wuppertal, 42097 Wuppertal, Germany,
$^{dd\dag}$Yarmouk University, Irbid 211-63, Jordan, and
$^{ll\dag}$On leave from J.~Stefan Institute, Ljubljana, Slovenia,
and with visitors to D0 from
$^{mm\ddag}$Augustana College, Sioux Falls, SD, USA,
$^{nn\ddag}$The University of Liverpool, Liverpool, UK,
$^{oo\ddag}$SLAC, Menlo Park, CA, USA,
$^{pp\ddag}$ICREA/IFAE, Barcelona, Spain,
$^{qq\ddag}$Centro de Investigacion en Computacion - IPN, Mexico City, Mexico,
$^{rr\ddag}$ECFM, Universidad Autonoma de Sinaloa, Culiac\'an, Mexico, and
$^{ss\ddag}$Universit{\"a}t Bern, Bern, Switzerland.
\end{small}

\end{document}